\documentclass[journal]{IEEEtran}
\ifCLASSINFOpdf
  \usepackage[pdftex]{graphicx}
   \graphicspath{{.}{../pdf/}{../jpeg/}}
   \DeclareGraphicsExtensions{.pdf,.jpeg,.png,.jpg}
\else
\fi
\usepackage[cmex10]{amsmath}
\usepackage{array}
\newcommand{\F}{{F}}
\newcommand{\Fd}{{F^\dag}}
\newcommand{\G}{{G}}

\newcommand{\Sa}{{S}}
\newcommand{\Sd}{{S^\dag}}

\newcommand{\W}{{W}}

\usepackage{enumerate}
\begin{document}

\title{Advances in Calibration and Imaging Techniques in Radio Interferometry}

\author{U.~Rau,
        S.~Bhatnagar,
        M.A.~Voronkov,
        and~T.~J.~Cornwell 
%\author{XXX-YYY,~\IEEEmembership{Member,~IEEE,}
%        XXX~YYY,~\IEEEmembership{Fellow,~IEEE,}
%        and~XXX~YYY,~\IEEEmembership{Life~Fellow,~IEEE}% <-this % stops a space
\thanks{U. Rau is with the Department of Physics, New Mexico Institute of Mining and Technology,
	Socorro, NM, USA;   
	National Radio Astronomy Observatory, Socorro, NM, USA; 
	CSIRO Australia Telescope National Facility, Marsfield, NSW, Australia,
		e-mail: rurvashi@aoc.nrao.edu.}% <-this % stops a space
\thanks{S. Bhatnagar is with the National Radio Astronomy Observatory, Socorro, NM, USA.}% <-this % stops a space
\thanks{M.A.Voronkov and T.J.Cornwell are with the CSIRO Australia Telescope National Facility, Marsfield, NSW, Australia.}% <-this % stops a space
%\thanks{(The National Radio Astronomy Observatory (NRAO) is a facility of the National Science Foundation operated under cooperative agreement by Associated Universities, Inc.)}%
\thanks{Manuscript received November 10, 2008; revised January 8, 2009.}}


%%\markboth{Journal of IEEE XXXX,~Vol.~6, No.~1, January~200X}%
\markboth{IEEE Special Issue on Advances in Radio Telescopes (in press)}%
{Rau \MakeLowercase{\textit{et al.}}: Advances in Calibration and Imaging Techniques in Radio Interferometry}



\maketitle

\begin{abstract}
%\boldmath
  This paper summarizes some of the major calibration and image
  reconstruction techniques used in radio interferometry and describes
  them in a common mathematical framework. The use of this framework
  has a number of benefits, ranging from clarification of the
  fundamentals, use of standard numerical optimization techniques, and
  generalization or specialization to new algorithms.
\end{abstract}

\begin{IEEEkeywords}
radio interferometry, calibration, imaging, algorithms, computing.
\end{IEEEkeywords}

% \fi
\IEEEpeerreviewmaketitle

%
%====================================================================
%
\section{Introduction}
%\IEEEPARstart{A}{perture} synthesis is an indirect imaging technique in which 
%the spatial Fourier transform of the image of an object is measured via its
%mutual coherence function.

\IEEEPARstart{T}{he} theory and practice of radio interferometry, including data
processing, is well-advanced and has been the subject of a graduate
level textbook \cite{THOMPSON_AND_MORAN}. This book is recommended for
the detailed descriptions of the fundamentals. In this paper, we aim
to summarize recent advances in the theory and practice of calibration
and imaging, arising from the work of several of the authors over the
past ten years. We draw upon a number of our papers, placing the
results in a common framework and nomenclature. We also present a
number of new insights and algorithms arising in recent work.

The last decade has seen a substantial growth in the number and
diversity of radio synthesis telescopes being constructed Examples
include the Expanded Very Large Array (EVLA, \cite{EVLA}, the Low Frequency Array (LOFAR,
\cite{LOFAR}), the Square Kilometre Array (SKA, \cite{SKA}), the Australian
Square Kilometre Array Pathfinder (ASKAP, \cite{ASKAP}) and
the Karoo Array Telescope (MeerKAT \cite{MEERKAT:URSI08}). These telescopes bring both
new science and new technical challenges. Prime amongst these
challenges are:
\begin{itemize}
\item Theory to describe new observing modalities and previously
  ignorable effects,
\item Algorithms to solve the resulting equations,
\item A required increase in algorithmic performance in terms of
  sensitivity and dynamic range,
\item A large increase (hundreds or thousands) in data volume,
\item The need for algorithms adapted to high performance computing,
  particularly the shift to highly parallel or concurrent processing.
\end{itemize}

The concept of a {\it measurement equation} is key to our work. Hamaker,
Bregman, and Sault \cite{HBS1} were particularly notable in
emphasizing the importance of a single equation to describe a
measurement process (as opposed to, say, a set of loosely related
equations). 

Section~\ref{Measurement Equation} describes the measurement equation
in radio interferometry. Section~\ref{Standard Algorithms} describes
the solution of the measurement equation as an optimization problem
and describes standard algorithms and methods used to solve it -
calibration of direction independent instrumental effects and imaging
using a point-source flux model.  Section~\ref{Sec:Direction Dependent
Imaging} describes recent advances in algorithms that account for
direction dependent instrumental effects during imaging. Section
\ref{Parameterized Deconvolution} describes recent advances in
deconvolution algorithms.
%All discussions are of numerical algorithms, described
%within a mathematical framework amenable to software implementation using 
%standard optimization libraries.
%Section \ref{Computing} deals with software
%design issues and high performance computing needs for current and
%next-generation radio telescopes.

%%\hfill mds
 
%%\hfill November 1, 2008

%%%%%%%%%%%%%%%%%%%%%%%%%%%%%%%%%%%%%%%% START MAIN TEXT
%\section{Prior Work - TJC}
%References to prior work and what this paper aims to do.
%
%====================================================================
%
\section{Measurement Equation in Radio Interferometry}
\label{Measurement Equation}
Aperture synthesis is an indirect imaging technique where the spatial 
Fourier transform of an image is measured via its mutual coherence function.
%
%For electromagnetic radiation from a spatially incoherent brightness distribution, 
%the mutual coherence function is defined as the time averaged cross correlation product of
%the total electric field measured at two aperture points with a time delay 
%between the measurements.
%It can be shown that if the source is in the far field, and if 
%the total time delay between aperture points is smaller than the coherence time given
%by the inverse bandwidth of the signal,
%the mutual coherence function becomes a (spatial autocorrelation) visibility function
%that measures the spatial Fourier transform of the brightness distribution.
%This is known as the Van-Cittert-Zernike theorem.
%
A radio interferometer \cite{NRAO_LECTURES} consists of a collection of spatially separated antennas.
%antennas spread across the ground.
The aperture plane of the interferometer 
is the plane perpendicular to the instantaneous direction from the array to a reference point 
on the sky $\vec{s}_0$ called the phase-reference center. 
A {\it baseline} $\vec{b}_{ij}$ is defined as 
the vector between the 3D locations of two antennas $i$ and $j$,
projected onto this aperture plane.
%In a co-ordinate system defined on this plane, 
%The baseline vector in units
%of wavelength is given by $(u,v,w)=(\vec{b}_{ij})/\lambda$ where
The components of $\vec{b}_{ij}$ are measured in units of wavelength $\lambda$ and denoted
as $u,v,w$ where $u,v$ are 2D spatial frequencies 
and $w$ describes the height of an antenna
relative to the plane of the array in the direction of $\vec{s}_0$.
%The mutual coherence function of the total electric field incident on the array
%(in an inertial reference frame) is given by
For electromagnetic radiation from a spatially incoherent brightness distribution, 
the mutual coherence function is defined as the time averaged cross correlation product of
the total electric field measured at two aperture points (antennas) with a time delay 
between the measurements, and is given by
\begin{equation}
\Gamma(\vec{b}) = \int \left\langle E(\vec{s},t) \cdot E^*(\vec{s},t-{\vec{b}\cdot \vec{s}}/{c})\right\rangle e^{{-2\pi i{\vec{b}\cdot\vec{s}}/\lambda}}d\Omega
\label{Eq:MUTUAL-COHERENCE}
\end{equation}
where $\vec{s} = \vec{s}_0 + \vec{\sigma}$ 
describes a point near the phase reference centre,
$E(\vec{s},t)$ is the complex amplitude of the radiation emanating from a 
source in the direction $\vec{s}$,
${\vec{b}\cdot \vec{s}}/{c} $ is the time difference %between the propagation delay 
between the incoming radiation collected at two antennas separated by $\vec{b}$,
and $d\Omega = d\vec{s}/R^2$ where $R$ is the distance between the source and 
the aperture plane.

Signals from all antennas are delay corrected by a common factor given by 
${\vec{b}\cdot \vec{s}_0}/{c}$, to steer the array towards $\vec{s}_0$. 
If the maximum remaining delay ${\vec{b}\cdot \vec{\sigma}}/{c} $ is smaller than
the signal coherence time, the
term in the angle brackets becomes the source autocorrelation function or the
three-dimensional source brightness distribution $I(l,m,n)$, where 
$l,m,n=\sqrt{1-l^2-m^2}$ are direction cosines describing $\vec{\sigma}$.
Eq.~\ref{Eq:MUTUAL-COHERENCE} becomes
\begin{equation}
%%V(u,v) = \int \int I(l,m) e^{-2\pi i (ul+vm)} dldm
V(u,v,w) = \int \frac{I(l,m,n)}{n} e^{-2\pi i (ul+vm+w(n-1))} dl dm
\label{Eq:VCZ}
\end{equation}
When the array is coplanar ($w\approx 0$), 
or the region of the sky being imaged may be assumed flat ($n\approx 1$), 
Eq.~\ref{Eq:VCZ} describes a 2D spatial Fourier transform
relation between the mutual coherence function and the source brightness.
This is the Van Cittert Zernike theorem \cite{THOMPSON_AND_MORAN} and forms the basis for
interferometric imaging.

To measure polarised radiation \cite{HBS1}, two nominally orthogonal components of the incident electric
field $\vec{E_i} = [E^p~E^q]_i^T$ are measured at each antenna $i$.
Four cross-correlation pairs (two cross-hand and two parallel-hand) are formed
per baseline as $\langle \vec{E_i} \otimes \vec{E^*_j} \rangle$. 
The resulting coherence vector is denoted as
$\vec{V_{ij}}=[V^{pp}~V^{pq}~V^{qp}~V^{qq}]_{ij}^T$.
The vector of images corresponding to the four correlations is 
$\vec{I}=[I^{pp}~I^{pq}~I^{qp}~I^{qq}]^T$ 
and is related to the standard Stokes vector of images by a 
linear transform.

The measured incoming radiation is modified by propagation effects and receiver
electronics.
Jones matrices describe this modulation for the electric field incident
at each orthogonal pair of feeds $\vec{E_i} = [E^p~E^q]_i^T$. 
Direction independent effects for antenna $i$ are described as $J_i^{vis} = [G D C]$,
a $2\times2$ matrix product of 
complex antenna gains ($G$), polarisation leakage ($D$) and feed configuration ($C$). 
Direction dependent effects are described by $J_i^{sky} = [E P F]$, a product of
antenna illumination patterns ($E$), parallactic angle effects ($P$) and tropospheric and
ionospheric effects and Faraday rotation ($F$).
The effect on each baseline $ij$ is described by the outer-product of these antenna-based
Jones matrices given by $K^{\{vis,sky\}}_{ij} = [J_i \otimes J_{j}^{\dag}]^{\{vis,sky\}}$, 
a $4 \times 4$ matrix. (In this paper, the $\dag$ superscript is used to
denote conjugate transpose or operator adjoint.)

The measurement equation \cite{NRAO_LECTURES} for one baseline (spatial frequency), one frequency
channel, and one integration timestep, is given by
\begin{equation}
\vec{V}^{obs}_{ij} = [K^{vis}_{ij}] \int [K^{sky}_{ij}] \vec{I}^{sky}(\vec{s})e^{-2\pi i{\vec{b}\cdot\vec{\sigma}/\lambda}}d\Omega
%%\vec{V}^{obs}_{ij} = [K^{vis}_{ij}]\int\hspace{-0.25cm} \int [K^{sky}_{ij}] \vec{I}^{sky}(\vec{s})e^{-2\pi i{\vec{b}\cdot\vec{\sigma}/\lambda}}d\Omega
\label{Eq:VIJ2}
\end{equation}
%where $K^{\{vis,sky\}}_{ij} = [J_i \otimes J_{j}^{\dag}]^{\{vis,sky\}} + n_{ij}$,
%is a $4 \times 4$ matrix and 
%is the sum of the outer-product of antenna based Jones matrices and 
%$n_{ij}$ represents random additive baseline-based noise called {\it closure noise}.
%In the presence of random additive baseline based noise called {\it closure noise},
%$K_{ij}$ cannot be exactly decomposed into antenna based terms.
% random additive baseline-based measurement 
%noise called {\it closure noise}.
All instrumental and propagation effects described by $K_{ij}$ 
need to be corrected during image reconstruction.

%and 
%$J=\left( \begin{array}{ll}
%g_p & d_{pq} \\
%d_{qp} & g_{q} \\
%\end{array}\right) $
%where $g^p$ is the complex gain of the $p$ feed, and $d^{pq}$ is the amount of 
%leakage from the $q$ feed into the $p$ feed, and  $\vec{E}^{obs} = J \vec{E}$
%The standard Stokes vector that
%describes the polarisation state of the EM radiation across the image is related to 
%$\vec{I}$ by a linear operator. % $S$ as $\vec{I} = [S] \vec{I}_{stokes}$.
%For example, Stokes I images can be made from $V^I_{ij} = V^{pp}_{ij} + V^{qq}_{ij}$. 

So far, we have dealt with the signals measured at only one baseline.
With $n_{ant}$ antennas, there are $n_{ant}(n_{ant}-1)/{2}$ baselines that
make simultaneous measurements at multiple spatial frequencies.
The spatial frequency plane can be further sampled by varying the positions of the
antennas with respect to the direction of the phase-reference center. 
For ground-based arrays, the Earth's rotation makes all projected baseline 
vectors $\vec{b}\cdot\vec{s}_0$ trace ellipses on the spatial frequency plane, slowly filling it up.
Measurements at multiple receiver frequencies also increase the sampling of the 
spatial-frequency plane.
Measurements must be made at sufficiently high time and frequency resolution, to prevent
smearing (averaging of visibility data) on the spatial frequency plane.
The result is generally a centrally dominated $uv$-plane sampling pattern with a hole in the
middle and tapered outer edges. This is the transfer function of the synthesis array and
is called the {\it uv-coverage} (see \cite{NRAO_LECTURES}). 

The complete measurement equation can be written in matrix notation to include the effect
of the uv-coverage.
Let $I^{sky}_{m\times 1}$ be a pixelated image of the sky and let $V^{obs}_{n\times 1}$
be a vector of $n$ visibilities. 
Let $S_{n\times m}$ be a projection operator that describes
the uv-coverage as a mapping of $m$ discrete spatial 
frequencies (pixels on a grid) to $n$ visibility samples (usually $n>m$).
Let $F_{m\times m}$ be the Fourier transform operator and
$c$ be the number of measured correlations (1, 2 or all 4 of $\{pp,pq,qp,qq\}$).
The measurement equation in block matrix form is
\begin{equation}
\vec{V}^{obs}_{cn\times 1} = [K^{vis}_{cn\times cn}] [S_{cn\times cm}] [F_{cm\times cm}] [K^{sky}_{cm\times cm}] \vec{I}^{sky}_{cm\times 1}
\label{Eq:VIJ3}
\end{equation}

Writing this completely in the spatial frequency domain,
\begin{equation}
\vec{V}^{obs}_{cn\times 1} = [K^{vis}_{cn\times cn}] [S_{cn\times cm}][G_{cm\times cm}]\vec{V}^{sky}_{cm\times 1}
\label{Eq:VIJ-FT}
\end{equation}
where $[G_{cm\times cm}] = [F_{cm\times cm}] [K^{sky}_{cm\times cm}] [\Fd_{cm\times cm}]$ is a
convolution operator\footnote
{
\label{FN:One}
A 1-D convolution operator is constructed as follows.
Consider $\vec{a} \star \vec{b}$. Let $[A]$ and $[B]$ be diagonal matrices constructed
from vectors $\vec{a}$ and $\vec{b}$ respectively. 
Then, $A \star B = \Fd (\F A)( \F B) = [\Fd A_F \F]B = [C][B]$.
Here, $diag(A_F)$ is the discrete Fourier transform (DFT) of $\vec{a}$, and $[C]$ is a
Toeplitz matrix, with each row containing a shifted version of $\vec{a}$.
Multiplication of $[C]$ with $\vec{b}$ implements the shift-multiply-add sequence
required for the process of convolution. 
Since $\F$ is unitary, the singular-value-decomposition of $[C]$ is given by
the Fourier transform, making it circulant. 
For a 2D convolution, $[\F]$ is the outer product of two 1-D DFT operators
and $[C]$ is block-circulant with circulant blocks.
} 
in the spatial frequency domain with $[F K^{sky}]$ as the convolution filter. 

All discussions that follow are of numerical algorithms, described
within a mathematical framework amenable to implementation using 
standard optimization software.

%For standard imaging not accounting
%for direction-dependent effects, $[G_{cm\times cm}]$ is the operator
%for re-sampling the data onto a regular grid.

%Section \ref{Standard Algorithms} describes the solution of the measurement equation for
%a single correlation ($pp$) with only direction-independent complex antenna gains $G_i$  
%($K_{ij}$ is scalar, $K^{sky}_{ij}=1$) and point sources.
%Section \ref{Sec:Direction Dependent Imaging} describes new algorithms to account for
%direction dependent calibration terms ($K^{sky}_{ij} \neq 1$) for a single correlation,
%non-coplanar baselines ($w \neq 0$) and a wide field of view ($n \neq 0$).
%Section \ref{Parameterized Deconvolution} describes imaging of extended 
%emission using multi-scale and multi-frequency source models.
%Section \ref{Full Stokes} describes full-Stokes calibration and imaging 
%that uses all cross-correlations $\{pp,pq,qp,qq\}$ with the full $K_{ij}^{\{vis,sky\}}$. 
%
%====================================================================
%
\section{Standard Calibration and Imaging}
\label{Standard Algorithms}
This section describes the solution of the measurement equation as a numerical
optimization problem. 
The measurement equation for a single correlation with no direction-dependent terms is given as
\begin{equation}
{V}^{obs}_{n\times 1} = [K^{vis}_{n\times n}] [\Sa_{n \times m} \F_{m\times m}] {I}^{sky}_{m\times 1}
\label{SIMPLEMEQN}
\end{equation}
%With only one correlation, $K^{vis}_{ij}=g^p_i g^{*p}_j$ is a scalar and $K^{vis}$ 
%is a diagonal matrix.
Consider only the $pp$ correlation product, and let the complex
gains per antenna $i$ be given by $[G_i]=g^p_i$. 
Then, $K^{vis}_{ij}=G_i \otimes G_j = g^p_i g^{*p}_j$
is a scalar and $[K^{vis}_{n\times n}]$ is a diagonal matrix.

The unknowns in Eq.~\ref{SIMPLEMEQN} are the sky brightness $I^{sky}$ and
the complex gain product for all visibilities $K^{vis}$.
A two-stage $\chi^2$ minimization 
process iterates between these two parameter subspaces and applies constraints
appropriate to the different physics involved. 
Calibration (section \ref{CALIBRATION}) is the process of computing and applying the 
inverse of $[K^{vis}]$.
Imaging (section \ref{IMAGING}) is the process of reconstructing the sky brightness $I^{sky}$
by removing the effect of the instrument's incomplete spatial frequency sampling.

\subsection{Calibration}\label{CALIBRATION}
The elements of $[K^{vis}]$ are first estimated from observations of a 
source whose structure is known a-priori ($V^{model}_{n\times 1}$)
by solving Eq.~\ref{Eq:VIJ-FT} in the form
\begin{equation}
V^{obs}_{n\times 1} = [K^{vis}_{n\times n}] V^{model}_{n\times 1}
\label{eqnSELFCAL}
\end{equation}
A weighted least squares solution \cite{SELFCAL_CORNWELL} 
of Eq.~\ref{eqnSELFCAL} is found 
by minimizing $\chi^2 = \sum_{ij} w_{ij} |V^{obs}_{ij} - g_i g_j^* V^{model}_{ij}|^2$ 
where $w_{ij}$ is a measured visibility weight (inverse of noise variance)
and $V^{model}_{n\times 1}$ provides $O(n_{ant}^2)$ constraints  
to uniquely factor the baseline-based $K^{vis}$ into $n_{ant}$ antenna-based complex 
gains.
%\footnote
%{
%For a single point-source at the phase centre, the rank one 
%matrix given by $X = \vec{g} \vec{g^{\dag}}$ is the covariance matrix of the signals 
%measured at each antenna and the computation of $\vec{g}$ is equivalent to the   
%eigen value decomposition of $X$ in the presence of noise.
%}.
In cases where the measurements at each baseline contain random additive noise
that cannot be factored into antenna-based terms ({\it closure noise}),
a baseline-based calibration is sometimes done to solve for 
the elements of $[K^{vis}]$ directly.

$[K^{vis}_{n\times n}]^{-1}$ is reconstructed from these solutions, and applied to the
observed visibilities to {\it correct} them.
\begin{equation}
V^{corr}_{n\times 1} = [K^{vis}_{n\times n}]^{-1} V^{obs}_{n\times 1}
\label{eqnCALIBRATION}
\end{equation}

To increase the signal to noise ratio of correlations going into the algorithm,
the visibility data are sometimes pre-averaged along data axes over which the solution is likely
to remain stable. 
%For example, bandpass calibration is usually done on time-averaged data 
%because the bandpass shape is usually stable across time. 
%Time-variable gain fluctuations are solved for during a second pass, 
%where the now calibrated bandpasses are averaged across channels 
%for each time-step. 

\subsection{Imaging}\label{IMAGING} 
Using Eqs.~\ref{SIMPLEMEQN} and \ref{eqnCALIBRATION}, the measurement equation after
calibration is 
\begin{equation}
[\Sa_{n\times m} \F_{m \times m}] I^{sky}_{m \times 1} = V^{corr}_{n \times 1}
\end{equation}
A weighted least squares estimate of $I^{sky}$
is found by solving the {\it Normal Equations}
\begin{equation}
[\Fd \Sd \W \Sa \F ] I^{sky}_{m\times 1} = [\Fd \Sd \W] V^{corr}_{n\times 1}
\label{NEQN}
\end{equation}
where $W_{n\times n}$ is a diagonal matrix of signal-to-noise based measurement weights
and $\Sd$ denotes the mapping of measured visibilities onto a spatial frequency grid.

The Hessian $[\Fd\Sd\W\Sa\F]$ on the LHS of Eq.~\ref{NEQN} describes the 
imaging properties of the instrument, and the RHS describes 
the raw image produced by direct Fourier inversion of the calibrated visibilities.
When $V^{corr}_{n\times 1} = \vec{{\bf 1}}_{n\times 1}$, the RHS gives the {\it impulse response function}
or {\it point spread function} of the instrument ($I^{psf}$), 
defined as the image produced when observing a point-source of unit brightness
at the phase center.
The Hessian is by construction, a circulant convolution operator
with a shifted version of $I^{psf}$ in each row.
Therefore, the {\it dirty image} produced by direct Fourier inversion of 
the measurements is the convolution of the true image $I^{sky}$ with the {\it PSF} of the instrument
and the Normal equations can be solved via a deconvolution.

Since $\Sa$ represents an incomplete sampling of spatial frequencies 
(column rank of $\Sa_{n\times m}$ is $<m$), the Hessian is singular.
Therefore although this convolution is a linear operation, the Hessian 
cannot be directly inverted to create a linear deconvolution operator.
Instead, an iterative Newton-Raphson approach is implemented as follows.
\begin{enumerate}[(a)]
\item Initialise the model image $I^m_{0}$ to zero or to a model that 
represents a-priori information about the true sky.
\item \label{step2}{\it Major Cycle} : Compute the $\bigtriangledown \chi^2$ (residual) image.
\begin{equation}
I^{res} = \left\{[\Fd\Sd\W][V^{corr}-[\Sa\F] I^{m}_{i})]\right\}
\end{equation}
The {\it forward} transform $V^{m}=[\Sa\F] I^{m}_{i}$ {\it predicts} 
visibilities that would be measured for the current sky model and
residuals are computed as $V^{res}=V^{corr}-V^{m}$.

The {\it reverse} transform $I^{res}=[\Fd\Sd\W] V^{res}$ computes an image 
from a set of visibilities. 
A {\it Preconditioning} scheme decides how best to weight the visibility data 
(see \ref{Preconditioning}) before {\it Gridding} them onto a regular grid of
spatial frequencies (see \ref{Gridding}) and Fourier inverting to give $I^{res}$.

\item {\it Minor Cycle} : Compute the update step by applying an operator $T$ to
the $\bigtriangledown \chi^2$ image. Update the model image.
\begin{equation}
I^{m}_{i+1} = I^{m}_{i} + T\left(I^{res},I^{psf}\right)
\label{DECONV_ITER}
\end{equation}
$T$ represents a non-linear deconvolution of the {\it PSF} from $I^{res}$
while filling-in unmeasured spatial frequencies (null space of the measurement matrix)
for a complete reconstruction of the image.
%\footnote
%{
%Following the standard calculation for the update step in a $\chi^2$ minimization,
%$T\left(I^{res},I^{psf}\right) = [\Fd\Sd\W\Sa\F]^{-1} I^{res}$.
%However, in our case since the Hessian is singular, this form of 
%$T$ is never explicitly computed.  
%}
%of the {\it PSF} from $I^{res}$ while filling in
%unmeasured spatial frequencies.
Section \ref{Deconvolution} describes
$T$ for several standard deconvolution algorithms.
\item Repeat from (\ref{step2}) until convergence is achieved ($I^{res}$ is noise-like)
or other termination criteria are satisfied ($T$ can no longer reliably 
extract any flux from $I^{res}$).
\item The final $I^m$ is {\it restored} by first smoothing it to 
the maximum angular resolution of the instrument to suppress artifacts
arising from unconstrained spatial frequencies beyond the measured range
and then adding in the final $I^{res}$ to preserve any undeconvolved flux.
\end{enumerate}

%\begin{equation}
%\triangle I^m_i = [\Fd\Sd\W\Sa\F]^{-1}\left\{[\Fd\Sd\W][V^{corr}-[\Sa\F] I^{m}_{i})]\right\}
%\end{equation}

%% Flag for 'new' stuff..
\subsubsection{Preconditioning}\label{Preconditioning}
The aim of preconditioning is to alter the shape of the {\it PSF} according to
whatever makes the Normal equations easier to solve. 
This is done by re-weighting
the data to tune the instrument's sensitivity to a particular type of source and
signal-to-noise ratio \cite{NRAO_LECTURES}.
%
%partially diagonalize the Hessian to make its inversion more 
%tractable during deconvolution. 
%Off-diagonal elements in the Hessian come from the finite width of the main 
%lobe of the PSF and the presence of far-out sidelobes. 
%Preconditioning is done by reweighting the data to modify the shape of the PSF according
%to whatever makes the Normal equations easier to solve. 
%
%Several weighting schemes are used to alter the shape of the {\it PSF} and
%tune the instruments sensitivity to a particular type of source and signal-to-noise.

The {\it Natural Weighting} scheme gives equal weight to all samples and 
preserves the instrument's peak sensitivity, making it ideal for the detection of low
signal-to-noise sources. However, the non-uniform sample density on the uv-grid
can give the {\it PSF} a wide main lobe and high sidelobes.
%% {\tt CHECK : does psf improve with more ants ?}.
{\it Uniform Weighting} gives equal weight to each measured spatial frequency irrespective
of sample density and this lowers its peak sensitivity.
The resulting PSF has a narrow main lobe and suppressed sidelobes across the
entire image and is best suited for sources with high signal-to-noise ratios 
to minimize sidelobe contamination between sources.
{\it Super-Uniform Weighting} gives a {\it PSF} with 
inner sidelobes suppressed as in Uniform weighting but far-out sidelobes 
closer to that with Natural weights. The peak sensitivity is also closer to Natural weighting.
{\it UV Tapering} suppresses high spatial frequencies and
tunes the sensitivity of the instrument to peak for scale sizes 
larger than the resolution element.
{\it Robust Weighting} \cite{DANS_THESIS} creates a {\it PSF} that smoothly varies between Natural and Uniform
weighting based on the signal-to-noise ratio of the measurements and a tunable parameter
that defines a noise threshold.
%{\it Wiener Filtering} is a robust linear deconvolution that weights spatial
%frequencies with high signal-to-noise data uniformly, and suppresses spatial frequencies with low
%signal-to-noise ratios. 
%It partially diagonalizes the Hessian by computing and applying its
%SVD inverse.

The final {\it imaging weights} are given as $\W^{im}=\W^{pc}\W$ where
$\W^{pc}$ are preconditioning weights and $\W$ are measurement-noise based weights.
The Hessian becomes a convolution operator with the preconditioned
{\it PSF} in each row ($I^{psf} = diag[\Fd\Sd\W^{im}]$).

%The final {\it imaging weights} are given as $\W_{i}=\W_{pc}\W_{\sigma}$ where
%$\W_{pc}$ are preconditioning weights and $\W_{\sigma}$ are natural weights.
%The Hessian becomes a convolution operator with the preconditioned
%$PSF = diag[\Fd\Sd\W_i]$ in each row.

%Let $\W^G_{\sigma} = \Sd\W_{\sigma}\Sa$ represent the gridded weights.
%\begin{equation}
%[\Fd\W_{pc}\F][\Fd\Sd\W_{\sigma}\Sa\F] = m [\Fd\W^G_{i}\F]
%\end{equation}
%where $\W_{pc}$ are the preconditioning weights and 
%$\W^G_{i}=\W_{pc}\W^G_{\sigma}$ are the final imaging weights.

\subsubsection{Gridding} \label{Gridding}
The measured visibilities sample the spatial frequency plane along elliptical tracks 
and need to be binned onto a regular grid of spatial frequencies so that
the FFT algorithm can be used for Fourier inversion. 
Gridding interpolation is done as a convolution \cite{NRAO_LECTURES}. 
Each weighted visibility is first multiplied
with a prolate spheroidal function $P_s$ centred at its true location. 
Then, values at the centres of all grid cells within a certain radius are read off.
$P_s$ acts as an anti-aliasing function. 
A {\it grid correction} is then done in the image domain to remove this multiplicative
image-domain effect.

Let $P_s$ be a diagonal matrix representing the prolate-spheroidal function.
$G^{gc}=[\F(\Fd P_s)\Fd]$ is the corresponding gridding convolution operator in the
spatial frequency domain, equivalent to multiplying the image domain by $I^{wt}_{gc}=[\Fd P_s]_{m\times m}$.
The normalized {\it dirty image} and {\it PSF} are computed as 
\begin{equation}
I^{\{dirty,psf\}}_{m\times 1} = w_{sum}^{-1}[I^{wt}_{gc}]^{-1}[\Fd\G^{gc}\Sd\W^{im}] V^{\{corr,1\}}_{n\times 1} 
%%		&=& w_{sum}^{-1}[I^{gc}]^{-1}[\Fd P_s][H] I^{\{sky,\delta\}}
%%I^{\{dirty,psf\}}_{m\times 1} &=& w_{sum}^{-1}[I^{gc}_{m\times m}]^{-1}[\Fd\G_{gc}\Sd\W^{im}]_{m\times n} V^{\{corr,1\}}_{n\times 1} \\
%%		&=& w_{sum}^{-1}[I^{gc}_{m\times m}]^{-1}[\F^{-1} P_s][H]_{m\times m} I^{\{sky,\delta\}}_{m\times 1}
\label{DIRTYANDPSF}
\end{equation}
where division by $w_{sum} = trace(W^{im})$ normalizes the peak of the {\it PSF} to unity.
%$H = [\Fd\Sd\W^{im}\Sa\F]$ is the Hessian matrix
%whose diagonal elements are $w_{sum}$ (peak of $I^{psf}$).
%The peak of $I^{\{psf\}}$ is therefore normalized to 1
Eq.~\ref{DIRTYANDPSF} describes the practical implementation of 
the {\it reverse transform} of the {\it Major Cycle} and 
$I^{dirty}$ is the initial $I^{res}$ used to start the iterations.

The model image $I^{model}$ obtained at the end of each {\it Minor Cycle} is
used in the {\it forward transform} as
\begin{equation}
V^{m}_{n\times 1} = [\Sa\G^{gc}\F][I^{wt}_{gc}]^{-1} I^{m}_{m\times 1}
\end{equation}
The calculation of these transforms involves traversals of the entire
set of visibility data making it computationally expensive. 
Deconvolution algorithms usually tailor the frequency of {\it Major} and
{\it Minor} cycles to perform trade-offs between performance, accuracy and
total number of iterations.

\subsubsection{Deconvolution}\label{Deconvolution}
%This section describes the Minor Cycle.
For the {\it Minor Cycle}, $I^{dirty}$ is assumed to be a perfect 
convolution of the {\it PSF} with the true sky brightness,
%%($I^{dirty} = I^{sky} \star I^{psf}$
where $I^{\{dirty,psf\}}$ are given by Eq.~\ref{DIRTYANDPSF}.
The operator $T$ in Eq.~\ref{DECONV_ITER} constructs
a model image $I^m$ via a deconvolution. 
%This is a non-linear process because it attempts to fill in 
%the null space of the measurement matrix for a complete reconstruction of the image.

The {\it CLEAN} algorithm forms the basis for most deconvolution algorithms used in
Radio Interferometry. 
%% For filled aperture, inv exists and a Gauss-Siedel iteration set will solve it.
%% For partially filled aperture, need to solve N.E. and this is a matched-filtering method.
%% For Clean, uniform weights, the two procedures become the same !
The peak of the residual image %$I^{res} = I^{dirty} - I^{m}\star I^{psf}$ 
gives the location and strength of a potential point source. 
The effect of the {\it PSF} is removed by subtracting
a scaled $I^{psf}$ from $I^{res}$ at the location of each point source and updating $I^{m}$
(Eq.~\ref{DECONV_ITER}).
Many such iterations of finding peaks and subtracting {\it PSF}s form the Minor Cycle. 

The following deconvolution algorithms model the sky in a pixel basis and are best
suited to isolated point sources whose amplitude is constant across the
observing bandwidth.
Deconvolution algorithms that produce multi-scale and multi-frequency source models are 
described in section \ref{Parameterized Deconvolution}.

In {\it Hogbom CLEAN} \cite{CLEAN}, the Minor cycle subtracts a scaled and shifted version of the full 
{\it PSF} to update the residual image for each point source. Only one Major cycle
is done. It is computationally efficient but susceptible to errors due to inappropriate
preconditioning.
{\it Clark CLEAN} \cite{CLARK_CLEAN} does a set of Hogbom Minor cycle iterations 
using a small patch of the {\it PSF}. A Major Cycle is performed 
when the brightest peak in the residual image is
below the first sidelobe level of the brightest source in $I^{res}$.
The residual image is then re-computed
as $I^{res} = [\Fd](\F I^{dirty} - \F I^{m})$ to eliminate aliasing errors.
{\it Cotton-Schwab CLEAN} \cite{CSCLEAN1984} is similar to the Clark algorithm, but computes the residual
as $I^{res} = [\Fd\Sd\W](V^{corr} - [\Sa\F]I^{m})$.
It is time consuming but relatively unaffected by inappropriate preconditioning and
gridding errors because it computes $\chi^2$ directly in the measurement domain.
It also allows highly accurate prediction of visibilities without pixelation errors.
The {\it Steer-Dewdney-Ito CLEAN} Minor Cycle finds the locations of  
sources by setting an amplitude threshold to select pixels. 
The combined set of pixels is then convolved with the PSF and subtracted out 
via a Clark Major Cycle. This algorithm is more suited to deconvolving extended emission.
{\it Maximum Entropy (MEM)} \cite{MEM} methods and {\it Non negative least squares (NNLS)} 
\cite{DANS_THESIS},\cite{Lawson:SLS74} are
pixel-based deconvolution algorithms that perform a rigorous constrained
optimization in a basis of pixel amplitudes. MEM solves a least squares
problem with a penalty
function based on image entropy, that biases the estimate of the true sky brightness
towards a known prior image.
NNLS deconvolution solves a least-squares problem with linear inequality range constraints for all
its parameters.

%
%====================================================================
%
\section{Calibration and Imaging with Direction Dependent Instrumental
Effects}\label{Sec:Direction Dependent Imaging} 

$K^{sky}_{ij}$ in Eq.~\ref{Eq:VIJ2} represents the effects of
direction dependent (DD) gains in the measurement from a single
interferometric baseline.  These DD gains can result from a number
of instrumental and atmospheric/ionospheric effects, 
are potentially different for each baseline, and can be  
a function of time, frequency, polarisation and direction.
In the simplest form
of this equation these dependencies can be ignored, making
$K^{sky}_{ij}$ purely multiplicative in the image domain.
Imaging can then proceed as described in section \ref{Standard Algorithms}
(correcting only for {\it direction independant} terms), 
with the final image being divided by an estimate of $K^{sky}$ to 
remove the multiplicative DD effects. 
%Most conventional
%imaging algorithms implicitly assume the simplified measurement equation.

In its general form, Eq.~\ref{Eq:VIJ-FT} in the presence of DD effects for a telescope
calibrated for $\left[K^{vis}\right]$ can be written as
\begin{equation}
{V}^{obs}_{n\times 1}= \left[S_{n\times m}\right]\left[G^{dd}_{m\times m}\right] {V}^{sky}_{m\times 1}
\label{Eq:VIJ-DD1}
%where~\left[G^{dd}\right]=\left[F K^{sky} F^\dag\right]
%\label{Eq:VIJ-DD2}
\end{equation}
Each row of $\left[ G^{dd}_{m\times m}\right]$ acts as a
visibility-plane filter (see footnote \ref{FN:One}) for the measurements from baseline $ij$,
and is given by $[G^{dd}_{ij}]_{1\times m}=diag( [F K^{sky}_{ij}]_{m\times m})$,
%
%
%Each row of $\left[ G^{dd}_{m\times m}\right]$ 
%acts as a visibility-plane filter
%%\footnote
%%{
%%Note $F K^{sky}_{ij}$ is the visibility-plane representation of the
%%image-plane multiplicative term $K^{sky}_{ij}$. 
%%}
%for the measurements from baseline $ij$ (see footnote \ref{FN:One}) and is
%given by
%\begin{equation}
%[G^{dd}_{ij}]_{1\times m}=diag\left( [F K^{sky}_{ij} F^\dag]_{m\times m}\right)
%\end{equation}
where $K^{sky}_{ij}$ is assumed to be {\it known} from a-priori information.
Note that $K^{sky}_{ij}$ can also be separated into antenna based
terms.  We will exploit this property in Section~\ref{Sec:Pointing
SelfCal} to devise efficient solvers to solve for parametrized
forms of $\left[G^{dd}\right]$ for {\it unknown} DD effects.

%As is clear from Eq.~\ref{Eq:VIJ2}, when $K^{sky}_{ij}$ is different
%for each baseline, its image-plane effects cannot be described by a
%single multiplicative term for the entire image.  Consequently, even
%if the value of this term is known for each baseline, correction for
%its effects cannot be factored out of the imaging process, as is done
%for the correction of {\it direction independent} terms (see
%Section~\ref{CALIBRATION}).
%
%When $K^{sky}_{ij}$ is different for each baseline (as in Eq.~\ref{Eq:VIJ2}),
%two methods of correcting for DD effects are possible.  One approach
%requiring direct evaluation of the integral in Eq.~\ref{Eq:VIJ2} for
%the forward transform during iterative image deconvolution is
%discussed briefly in section~\ref{Sec:OBIT_SQUINT}.

%Equations \ref{Eq:VIJ-FT} and \ref{Eq:VIJ-DD1} suggests a second
%method based on FFT-based forward and reverse transforms to account
Equations \ref{Eq:VIJ-FT} and \ref{Eq:VIJ-DD1} suggest the use
of FFT-based forward and reverse transforms to account
for DD effects using an appropriately constructed $G^{dd}$ operator.
Data prediction can incorporate DD effects by using $G^{dd}$ as
the operator for re-sampling data from a regular grid (FFT of the
model image) at points given by the operator $S$. 
%A FFT based reverse transform which corrects for the DD effects is also
%possible by using
%$G^\dag_{gc}G^\dag_{dd}$ 
The reverse transform can correct for DD effects by using 
the conjugate transpose of $G^{dd}$ along with the standard
anti-aliasing operator $G^{gc}$ for gridding the data (see
Section~\ref{Gridding}). For such a transform to efficiently
correct for DD effects, the $G^{dd}_{ij}$ filter must satisfy two
properties: (1) it should have a finite support size (i.e.,
corresponding $K^{sky}$ should be band-limited), and (2) it should be
a unitary operator (or approximately so).  Effects of the W-term and
antenna primary beam patterns are two examples of DD effects, 
whose operators have these desirable properties.

% as:
%\begin{eqnarray}
%{V}^{Grid} &=& \left[G^{dd^\dag}\right] V^{Obs}\\
%           &=& \left[G^{dd^\dag}\right] \left[S\right]
%\left[G^{DD}\right] {V}^{sky}
%\label{Eq:REVERSE-TRANSFORM}
%\end{eqnarray}

A generalized version of Eq.~\ref{DIRTYANDPSF} including the DD
effects can be written as
\begin{equation}
I^{\{dirty,psf\}} = [I^{wt}_{dd}]^{-1} [I^{wt}_{gc}]^{-1}[\Fd\G^{gc}\G^{dd^\dag}\Sd\W^{im}] V^{\{corr,1\}} 
%		&=& [I^{wt}_{dd}]^{-1} [I^{wt}_{gc}]^{-1}[\Fd\G_{gc}\Gd_{dd}\F][H][\Fd\G_{dd}\F] I^{\{sky\}}_{m\times 1} \\
%		= [I^{wt}_{dd}]^{-1} [I^{wt}_{gc}]^{-1}[\Fd P_s][\K^{sky^\dag}][H] [\K^{sky}] I^{\{sky\}}_{m\times 1}
\label{Eq:PBDIRTY}
\end{equation}
where
\begin{eqnarray}
\label{Eq:PBWEIGHT}
{I^{wt}_{dd}} &=& [\Fd\G^{dd^\dag} \W^{im} \G^{dd}\F ] \\
\label{Eq:GRIDCORRWEIGHT}
{I^{wt}_{gc}} &=& [F^\dag P_s]
\end{eqnarray}
In the absence of DD effects, $G^{dd}$ is an identity matrix, 
$I^{wt}_{dd}=w_{sum} [1_{m\times m}]$ and Eq.~\ref{Eq:PBDIRTY} reduces to
Eq.~\ref{DIRTYANDPSF}.  $I^{wt}_{gc}$ is the same as the {\it grid
correction} mentioned in section~\ref{Gridding} to correct for the
image plane effects of the anti-aliasing operator $P_s$. 

Three special cases are discussed in the following sections.
\begin{enumerate}
\item When $G^{dd^\dag} G^{dd}$ is an identity matrix, $I^{wt}_{dd}$ is
still $w_{sum}$ and Eq.~\ref{Eq:PBDIRTY} can be used to generate
$I^{\{dirty\}}$ free of the relevant DD effects.  The effect of the
W-term discussed in section~\ref{Sec:WProjection} corresponds to
this case.

\item When $G^{dd^\dag}_{ij} G^{dd}_{ij}$ is a time dependent
function, $I^{wt}_{dd}=w_{sum}\left[K^{sky^\dag}K^{sky}\right]$.  DD
effects due to time varying antenna primary beams represent an example of
this case, as is discussed in section~\ref{Sec:Primary Beam
Correction}.

\item  Mosaic imaging or single pointing imaging with heterogeneous
antenna arrays corresponds to case where $G^{dd^\dag}_{ij}
G^{dd}_{ij}$ is not the same for all $i$ and $j$.  This is discussed
in section~\ref{Sec:Mosaicking}.
\end{enumerate}
%$\left[K^{sky}\right]$ for the antenna far field power pattern is
%band-limited and for most practical antennas is fundamentally
%non-circular.  Rotation of such a far field pattern with respect to
%the sky for El-Az mount antennas, random pointing errors heterogeneous
%antenna arrays (e.g. CARMA and ALMA) make $K^{sky}_{ij}$ variable with
%time and different for each baseline.  Similarly, the W-term in
%Eq.~\ref{Eq:VIJ2} ($w_{ij}\left[\sqrt{1-l^2 - m^2}-1\right]$) is not
%the same for all baselines and can also be represented by a
%$K^{sky}_{ij}$.
%\begin{itemize}
%\item Direction dependant effects cannot be factored out into a separate 
%calibration process. Need to combine with imaging.
%\item They appear as a gridding convolution function.\\
%Measurement Equation : $[\Sa \G \F] I^{sky}  = V^{corr} $\\
%Normal Equations ($[\Fd \Gd \Sd \W \Sa \G \F] I^{sky} = [\Fd \Gd \Sd \W] V^{corr}$).
%\item For each of the following, 
%describe the structure of $G$, the computation of the dirty image, psf,
%weight image and then refer to standard imaging algorithms.
%\end{itemize}
%
%====================================================================
%
\subsection {Correction for the W-term}
\label{Sec:WProjection}

The W-term is related to the fact that
Eq.~\ref{Eq:MUTUAL-COHERENCE} holds for coherence between the E-field
measured at two points on a common constant phase front of the
incident radiation \cite{BORN_AND_WOLF}. This is true only when the
array is coplanar, and the 
source being tracked is at the local zenith \cite{WFI_LECTURES2}.
Therefore in general, the image and visibility planes 
are not related by a 2D Fourier transform. The  
use of the 2D FFT for imaging wide-fields, results in a PSF which is no longer 
shift-invariant, making standard deconvolution algorithms unsuitable. 
However, if a Fresnel diffraction kernel is used as a propagator \cite{GOODMAN}
to compute the E-field measured at one of the antennas of each baseline, 
the 2D Fourier relation can be preserved. 
%Using Fresnel diffraction
%theory it can be shown that the E-field measured at one of the
%antennas of a baseline can be computed by propagating the E-field at
%an appropriate location on the constant phase front using the Fresnel
%propagator \cite{GOODMAN} and that this propagator is equal to the
This propagator is equal to the
Fourier transform of the W-term in Eq.~\ref{Eq:VCZ} ($e^{\iota w
\sqrt{1-l^2-m^2}}$).  
%Standard deconvolution algorithms are therefore not
%suitable.  
%A number of algorithms to correct for the effects of the
%W-term have been suggested.  Two most commonly used algorithms are
%described below.
Two algorithms commonly used to correct for the effects of the
W-term are described below.
%
%====================================================================
%
\subsubsection{Faceting algorithms}
The effect of the W-term is small close to the phase tracking center.
%(typically the same as the image center). 
This property is
exploited by algorithms which divide the field of view into
a number of facets.  Images are made by either projecting the facet
images onto the local tangent plane (image-plane faceting \cite{TIM_N_RICK_1992})
and using the appropriate PSF for the deconvolution of individual
facet images, or by projecting the $(u,v)$ for each facet onto a single
tangent plane in the gridding step required for an FFT-based reverse
transform \cite{UVFACETING}.  
This latter method produces a single flat image and has
several run-time and imaging performance benefits \cite{W_Projection_IEEE}.
%transform (see \cite{UVFACETING} for the details about co-ordinate
%projection).  This latter method produces a single flat image and has
%several run-time and imaging performance benefits (see
%\cite{W_Projection_IEEE} for a discussion).

%Image-plane facting algorith and a fully deconvolved image is independently made using the
%appropriate PSF for each facet.  Facting can however also be
%implemented in the visibility domain. The final deconvolved image is made
%by projecting these facet images to a common tangent plane image
%\cite{TIM_N_RICK_1992}.

%Number of facets required is given by
%\begin{equation}
%\label{Eq:NFACETS}
%N_f = \dfrac{\pi \Theta \sigma_w }{\sqrt{32 \delta A}}
%\end{equation}
%where $\Theta$ is the field of view in radians, $\sigma_w$ is the
%dispersion in the values of $w$ and $\delta A$ is the maximum
%tolerable errors in the amplitude. The number of floating point
%operations required is $O(N_f^2 + N_S^2)$ where $N_{S}$ is
%the support size of $G^{dd}_{ij}$ in pixels.
%
%====================================================================
%
\subsubsection{W-Projection algorithm}
In Eq.~\ref{Eq:VIJ-DD1}, the operator $\left[G^{dd}\right]$ can be
used to account for the W-term by choosing  
$K^{Sky}_{ij} = e^{w_{ij}\left(\sqrt{1-l^2-m^2}-1\right)}$.
%When $K^{Sky}_{ij} = e^{w_{ij}\left(\sqrt{1-l^2-m^2}-1\right)}$ in
%Eq.~\ref{Eq:VIJ-DD1}, the resulting operator $\left[G^{dd}\right]$ can
%be used to account for the W-term.  
This W-term operator $G^{dd}$ is
strictly unitary (by construction) and has a finite support (due to
the anti-aliasing operator $G^{gc}$). It will therefore 
correct for the W-term during image deconvolution \cite{W_PROJECTION,W_Projection_IEEE}.
%uses the FFT based fast forward
%and reverse transforms based on such a $G^{dd}$ operator
%%(Eqs.~\ref{Eq:VIJ-DD1} and \ref{Eq:REVERSE-TRANSFORM}) 
%to correct for the W-term during image deconvolution.
Conservatively speaking the W-Projection algorithm is about an order of
magnitude faster than faceting, and for the same amount of computing time can deliver
higher dynamic range images \cite{W_Projection_IEEE}.
%(see
%\cite{W_Projection_IEEE} for details and results of imaging with
%image-plane faceting and W-Projection algorithms).
%
%====================================================================
%
\subsection{Correction for Primary Beam}
\label{Sec:Primary Beam Correction}
%
%====================================================================
%
%Under the assumption that all antennas are identical and stable in
%time, the effects of antenna primary beam can be described by a single
%multiplicative term in the image domain.  Without loss of generality
%or imaging performance, the effects of primary beam can be ignored
%during imaging.  The resulting final image is tapered by the square of
%the primary beam pattern.  If a primary beam corrected image is
%required, this final image can be divided by a measured primary beam
%(or a reasonable model of it).  This is typically done in existing
%software. 
With the increased instantaneous sensitivities of next generation
telescopes and long integrations required for high dynamic range
imaging, antennas can neither be considered identical nor stable as a
function of time.  Therefore, next generation
imaging algorithms need to include corrections for the effects of 
time-varying antenna primary beams \cite{AWProjection_Memo,AWProjection}.
Algorithms to correct for these effects can be broadly
classified into two categories, namely corrections in the image plane
versus corrections in the Fourier plane.

\subsubsection{Image plane correction}
\label{Sec:OBIT_SQUINT}
When $K^{sky}_{ij}$ is different for each baseline,
one approach for correcting DD effects is the
direct evaluation of the integral in Eq.~\ref{Eq:VIJ2} for
the forward and reverse transforms during iterative image deconvolution
%Effects of non-identical primary beams can be incorporated in the
%forward and reverse transform for imaging by direct evaluation of the
%integral in Eq.~\ref{Eq:VIJ2}
\cite{USON_N_COTTON2008}.
%The computing load for
%forward and reverse transforms is $O(N_p \times N_d$) where for
%forward transform, $N_p$ is the number of pixels in the image with
%significant flux, $N_d$ is the number of visibility data points. $N_p$
%is the total number of pixels in the image for reverse transforms.
%Note that the computing load of each iteration of an iterative
%deconvolution algorithm is equivalent to one forward and one reverse
%transform.  Typically $N_d\sim10^6-10^9$, $N_p\sim10^6-10^7$ and
%algorithms take $\sim10-20$ iterations to converge.  The run-time,
%including the data i/o time (for data sizes in the range
%$1-100$~Tera~Bytes) can be prohibitive.  
The resulting run-time load for realistic data sizes can however be
prohibitive.  To reduce the compute load some-what, an FFT based reverse
transform (section~\ref{Gridding}) is used, but this
requires making assumptions about the variability 
of either the sky emission or the antenna power pattern.
%This is strictly unknown apriori
%making the usability of this approach severely limited.
%
%====================================================================
%
\subsubsection{Fourier plane correction -- The A-Projection algorithm}
\label{Sec:AProjection}

The visibility-plane filter describing the effects of the antenna
primary beams is the auto-correlation of the antenna aperture
illumination function.  For a finite sized antenna, this clearly has a
finite support in the Fourier domain.  However the resulting {\it
effective} operator ($\F G^{dd}/\sqrt{I^{wt}_{dd}}$) is only
approximately unitary \cite{AWProjection} .
%Consequently while an
%accurate FFT based forward transform that incorporates the effects of
%the primary beam can be implemented (Eq.~\ref{Eq:VIJ-FT}), the reverse
%transform is only approximate. 
%However, as described in
%Section~\ref{IMAGING}, the reverse transform in iterative
%deconvolution algorithms is used only to compute the approximate
%update direction (the residual image).  This is not different from the
%situation in most non-linear minimization problems involving data with
%noise.  As in the case of the W-Projection algorithm
The A-Projection algorithm uses accurate forward and approximate
reverse transforms based on the primary beam operator to correct for
time-variable primary beam effects (see \cite{AWProjection} for
details and an example of its application to full-beam imaging with the
VLA).
%The right- and left-circular
%beams for the VLA antennas are squinted with respect to each other due
%to antenna optics. This results in a strong instrumental Stokes-V
%signal which varies across the field of view and with time.  Using the
%AProjection algorithm, full beam thermal noise limited Stokes-I and
%Stokes-V imaging capability was demonstrated.  The primary beam
%operator was computed using a theoretical model for the antenna
%illumination pattern including feed and feed-legs blocking as well as
%antenna pointing errors and rotation of the antenna with parallactic
%angle with respect to the sky.  
Apart from the initial setup time required to compute the antenna
aperture function, the run time performance of this algorithm, when
imaging the entire field of view up to the first side lobe of the
antenna power pattern, is equivalent to that of standard image
deconvolution algorithms using a gridding convolution function with a support
size $\sim30\%$ larger in linear extent.

%All rows in $G$ are the same and it is a pure convolution operator.
%%Measurement Equation : $[\Sa \G \F] I^{sky} = [\Sa \F][\Pb] I^{sky} = V^{corr} $\\
%%Normal Equations ($[\Pb][\Fd \Sd \W \Sa \F][\Pb] I^{sky} = [\Pb][\Fd \Sd \W] V^{corr}$).
%
%====================================================================
%
\subsubsection{Pointing Self-Calibration}
\label{Sec:Pointing SelfCal}
Antenna pointing errors make $K^{sky}_{ij}$ (and the resulting $G^{dd}_{ij}$)
different for each baseline. 
When $K^{sky}$ represents effects of antenna primary beams, $K^{sky}_{ij}$ can be decomposed
%When $K^{sky}$ represents effects of antenna primary beams, antenna
%pointing errors make $K^{sky}_{ij}$ and the resulting $G^{dd}_{ij}$
%different for each baseline.  $K^{sky}_{ij}$ however can be decomposed
into two antenna based terms as $J^{sky}_i \otimes J^{sky^\dag}_j$,
each parametrized for  pointing errors, which 
can be recovered by solving the resulting parametrized
%each parametrized for antenna pointing errors. These antenna pointing errors
%can then in principle be recovered by solving the resulting parametrized
measurement equation.  However, iterative solvers using $K^{sky}_{ij}$ to
represent pointing errors necessarily require evaluation of the
integral in Eq.~\ref{Eq:VIJ2} in each iteration and have
proved to be impractically slow.

%Since $G^{dd}_{ij}$ parametrized for pointing errors can be
%efficiently computed, an alternate approach is to solve for the
%antenna pointing errors in the visibility domain.  Given a model for
An alternate approach is to solve for antenna pointing errors in 
the visibility domain, where it is efficient to compute $G^{dd}_{ij}$
parametrized by pointing errors.
Given a model for the sky, the Pointing SelfCal algorithm
\cite{POINTING_SELFCAL} iteratively solves for these pointing errors.
%using such a visibility domain parametrization.  
%An efficient implementation is
%possible since the fast forward transform described in
%section~\ref{Sec:AProjection} to compute the residuals and
%derivative in an iterative $\chi^2$ minimization algorithm.  Antenna
%pointing errors can then be included as part of $G^{dd}_{ij}$ used in
%the imaging deconvolution algorithm described in
%section~\ref{Sec:AProjection} to correct for the effects of pointing
%errors.
This algorithm can be efficiently implemented using the forward and reverse transforms described in
section~\ref{Sec:AProjection}. The effects of pointing errors can also be corrected 
along with other direction-dependent effects, as part of an iterative image deconvolution.

%% Apart from the obvious advantage of a run-time efficient algorithm,
%% such parameterization of DD effects has a more fundamental advantage;
%% since one is solving for few {\it antenna based} parameters, the solutions are
%% sensitive to all the flux in the entire field of view.  Other
%% approaches where one solves for an independent complex gain towards
%% multiple direction (towards strong compact sources) suffers from two
%% important fundamental deficiencies: (1) the solution for each
%% independent complex gain is sensitive only to the flux from small part
%% of the field of view, and (2) requires larger number of degrees of
%% freedom (DoF).  As a result, the signal-to-noise ratio per DoF can be
%% significantly smaller.

\subsection{Mosaicing}
\label{Sec:Mosaicking}
Mosaicing observations consist of a number of independent pointings 
covering a large field of view with an adequate sampling. 
Instruments with focal plane arrays can be considered to
observe a number of mosaic pointings in parallel, while traditional instruments observe
only one pointing at a time. Mosaicing observations can be treated in a natural way using
the formalism of Eqs.~\ref{Eq:PBDIRTY}-\ref{Eq:GRIDCORRWEIGHT}. 
Every pointing of the mosaic corresponds to a separate $G^{dd}$ and $I^{wt}_{dd}$. 
The difference  may be as little as the pointing direction (i.e. a translation of $I^{wt}_{dd}$ and 
phase gradient for $G^{dd}$), although  more substantial changes are possible 
(e.g. for  inhomogeneous arrays). Also, in the presence of noise Eq.~\ref{Eq:PBDIRTY} 
does not adequately constrain the dirty image in those parts of the sky where
the weight $I^{wt}_{dd}$ is low.
%Therefore, a combination of pointings with different centres is required to cover a larger area.
%Eq.~\ref{Eq:PBDIRTY} gives the product $I^{dirty}[I^{wt}_{dd}]_k$ for every pointing $k$, 
%forming a linear system of equations with respect to $I^{dirty}$. The solution is a generalization of 
%Eq.~\ref{Eq:PBDIRTY}  and is known as a linear mosaicing solution
The solution is a generalization of Eq.~\ref{Eq:PBDIRTY} where 
the product $[I^{wt}_{dd}]_kI^{dirty}$ for every pointing $k$ is combined to form a
linear system of equations. This is known as linear mosaicing.
\begin{equation}
%I^{\{dirty\}} = [I^{wt}_{dd}]^{-1} \sum\limits_{k}[I^{wt}_{gc}]^{-1} [\Fd\G^{gc}\G^{dd^\dag}_{k}\Sd\W_{im,k}] {V^{\{corr\}}_{k}}
I^{\{ dirty,\atop psf \}} = [I^{wt}_{dd}]^{-1} \sum\limits_{k}[I^{wt}_{gc}]^{-1} [\Fd\G^{gc}\G^{dd^\dag}_{k}\Sd\W^{im}_{k}] {V^{\{corr,1\}}_{k}}
\label{LINMOS}
\end{equation}
%where the weight is given by the same equation (\ref{Eq:PBWEIGHT}) as before with the only
%difference that index $k$ spans now the different mosaic pointings as well, and $G_{dd}$ is, in 
%general, $k$-dependent.
where the weight is given by a similar generalization of Eq.~\ref{Eq:PBWEIGHT}.
\begin{equation}
I^{wt}_{dd} = \sum\limits_{k} [\Fd\G^{dd^\dag}_{k} \W^{im}_{k} \G^{dd}_k\F]
\end{equation}

%The same approach applies to the {\it PSF}, which is  calculated as a response
%to a point source located at the centre of the mosaic (or any other location, 
%but the same for all pointings). 
%It is given by a generalization of (\ref{DIRTYANDPSF}) with a 
%summation over all pointings.
%Strictly speaking, the PSF obtained this way is
%valid only for one particular location. 
Strictly speaking, the PSF calculated as a response
to a point source located at the centre of the mosaic (or any other location;
but same for all pointings), is valid only for one particular location.
For any other direction in the field of view the contributions 
of individual pointings are different, causing a different  response. 
Therefore, the deconvolution performed in the minor cycle is always an approximate
operation and a number of major cycles is usually required.
However, this fact allows one to optimize the PSF calculation by taking into 
account only one pointing 
which contributes the most to Eq.~\ref{LINMOS} (e.g. the closest pointing to the centre
of the mosaic). Another way is to use a representative pointing and apply a phase shift to the 
convolution operator $G$ to centre the primary beam 
(i.e. to remove the offset of this pointing with respect to the mosaic centre).   

This approach to mosaicing is a form of a joint deconvolution, because the data from all
pointings are combined before the deconvolution takes place. 
It was shown to be superior to independent deconvolution where the final image is computed as a weighted sum 
of deconvolved sub-images corresponding to individual pointings of the mosaic  \cite{Cornwell88}. 
%However, the latter is often used, which implies two major approximations. First, each sub-image  
%obtained ignoring the DD effects should be a product of some average primary beam and the true brightness distribution. Second, the deconvolution operation can be reordered with the linear 
%mosaicing equation (strictly speaking it is not true because deconvolution is a non-linear operation).
%It is worth noting, that the general formalism presented in this paper does not justify the independent deconvolution.

%\subsection{Ionospheric Calibration}
%\subsection{Peeling}

%Curse of dimensionality \cite{BELLMAN}.
%Similar to eigen value decomposition assuming an orthogonal vector
%space (i.e. well separated compact sources and PSF without significant
%sidelobes).
%
%====================================================================
%
\section{Imaging Algorithms with advanced image parameterisations}\label{Parameterized Deconvolution}
So far, the discussions in this paper have focused on the calibration and imaging of
visibilities from one polarisation pair, the use of a pixel basis to
parameterize the sky brightness distribution, and the assumption that source structure is 
constant across the entire bandwidth of data being imaged.
In this section, we relax these assumptions and describe how standard methods can
be augmented to handle the added complexity of the increased dimensionality 
of the parameter space.

\subsection{Multi-Scale CLEAN Deconvolution}
Images of astrophysical objects tend to
%be a combination of point-like sources that
%are smaller than the instruments angular resolution, and extended emission 
show complex structure at different spatial scales.
%The incomplete uv-coverage of an interferometer 
%Unmeasured spatial frequencies fall into the null space of the measurement matrix 
%and reconstructed visibilities are largely unconstrained. 
%Weak constraints come from smoothness requirements in the spatial-frequency domain
%imposed during gridding convolution and image-domain support constraints.
%
%The strongest constraint comes from the choice
%of image parameterization, especially if it can describe all the structure in the
%source by a minimal number of parameters or basis functions.
The use of a pixel-basis for deconvolution is ideal for fields of isolated
point-like sources that are smaller than the instrument's angular resolution,
but tends to break extended emission into a collection of compact sources, 
   which is often inaccurate.
A better choice is to parameterize the image 
in a scale-sensitive basis that spans the full range of scale
sizes measured by the instrument. This provides a strong constraint on the
reconstruction of visibilities in the null space of the measurement matrix.
Also, since spatial correlation length fundamentally separates signal from noise,
scale-sensitive deconvolution algorithms generally give more noise-like 
residuals \cite{Asp_Clean}.

The Minor Cycle of the {\it Multi-Scale CLEAN} algorithm \cite{MSCLEAN}
parameterizes the image into a collection of inverted
tapered paraboloids ($h_k, k=1:n_{scales}$) 
whose widths are chosen from a predefined list.
{\it PSF}s and {\it dirty} images corresponding to each spatial scale are
calculated by smoothing $I^{\{dirty,psf\}}$ from Eq.~\ref{DIRTYANDPSF} by each $h_k$. 
Each iteration $i$ of the Minor cycle follows a matched-filtering technique 
where the location, amplitude and scale of each new component 
is chosen from ${max}\{I^{res}\star h_k\}$ ($\star$ denotes convolution)
and the update step 
%as $I^{res}_{i+1}\star h_k = I^{res}_i\star h_k - ({PSF}\star h_c \star {PSF} \star h_k)$
%where $c$ is the chosen scale size, 
accounts for the non-orthogonality of the different basis functions $h_k$.
MS-CLEAN works very well for complicated spatial structure but its performance
is limited by working with a discrete set of scale sizes, and
the fact that if an inappropriate component is chosen it takes the addition of
many more components to correct it. Typically, $n_{scales} \approx 8$ for a source
with complex spatial structure.
{\it Multi-Resolution CLEAN} \cite{MRCLEAN} performs a series of Hogbom Minor Cycles 
at different angular resolutions beginning at the lowest resolution to collect
all extended emission and progressing to higher resolutions. 
{\it PSFs} and residual images at different resolutions are made by varying the
image pixel sizes during gridding. 
Its limitations are similar to MS-CLEAN, in that there is no way to undo a 
component selection in case a better option becomes available later in the
iterations, and is less robust since it searches for components one scale
size at a time.
The {\it ASP CLEAN}  \cite{Asp_Clean} algorithm parameterizes the sky brightness
distribution into a collection of
Gaussians and does a formal constrained optimization on their parameters.
In the Major Cycle, visibilities are predicted analytically with high accuracy.
%from the expressions of the Fourier transforms of the Gaussians.
In the Minor Cycle, the location of a flux component is chosen from the peak
residual, and the parameters of the largest Gaussian that fits the image at 
that location are found. The minimization proceeds over subspaces consisting of 
sets of localized Gaussians whose parameters are varied together. 
This prevents errors due to inappropriate fits from propagating very far into the iterations.
The computing costs and runtimes of each Minor Cycle iteration of {\it MS-CLEAN} 
and {\it ASP-CLEAN} are a few times worse than {\it Hogbom-CLEAN}.
However, they parameterize the sky brightness more
physically and convergence is achieved in far fewer iterations.

\subsection{Multi-Frequency Synthesis Imaging}
The uv-coverage of a synthesis array can be greatly improved by using
the fact that 
visibilities measured at different receiver frequencies correspond
to different spatial frequencies.
%Traditionally, data from wide-bandwidth receivers is split into spectral channels and
%imaged separately to avoid {\it bandwidth smearing}.
%, a result of inaccurate mapping of visibilities onto spatial frequencies.
{\it Multi Frequency Synthesis} (MFS) is the process of 
combining data from multiple spectral channels onto the same spatial-frequency grid
during imaging to take advantage of the increased uv-coverage and imaging sensitivity.
As long as the 
sky brightness is the same across the total measured bandwidth, standard
imaging and deconvolution algorithms can be used along with MFS.
If the sky brightness varies across the observing bandwidth, the narrow-band
(or monochromaticity) requirement of aperture synthesis breaks down and 
the Fourier relation in the Van Cittert Zernike theorem does not hold. 
The following algorithms fold a frequency dependence of the image sky model 
into the measurement equation to handle this problem in the {\it Minor Cycle}.

{\it MF-CLEAN} \cite{MFCLEAN} is a matched filtering technique based on {\it spectral PSFs} 
that describe the instrument's responses to point sources with spectra given by
Taylor series functions (see Eqs.~\ref{MFDD},\ref{SpWts}). 
Source spectra ($I(\nu)$) are modeled as a power law and 
a first order Taylor expansion of $I(\nu)$ is combined with the regular 
imaging equation to describe the dirty image as a sum of convolutions
given by $I^{dirty} = \sum_t I^{dirty}_{t} = \sum_{t} C_{t} I^{m}_{t} \star I^{psf}_{t}$, where 
$I^{psf}_{t}$ are the {\it spectral PSFs} for $t=\{0,1\}$.
%\begin{equation}
%I^{dirty} = \sum_{t} C_{t} I^{m}_{t} \star \sum_{\nu} \left((\nu-\nu_0)/\nu_0)\right)^t I^{psf}_{\nu}
%\end{equation}
The deconvolution Minor Cycle simultaneously solves for $C_0$ and $C_1$ 
%solves for $C_0=I(\nu_0)$ and $C_1=\alpha I(\nu_0)$
for each component added to the model image.
%and $\alpha$ is calculated at the end of the cycle. 
This algorithm uses a pixel basis and is most
suited for point sources with pure power-law spectra with a weak frequency dependence.
%($\alpha=constant$)
{\it MS-MF-CLEAN} \cite{MSMFCLEAN} is a multi-scale multi-frequency deconvolution 
algorithm that extends MF-CLEAN to work with the instrument's
response to a polynomial spectrum ($n^{th}$ order Taylor series) 
at multiple spatial scales. 
%\begin{equation}
%I^{dirty} = \sum_{t,s} C_{t,s} I^{m}_{t,s} \star h_{s} \star \sum_{\nu} \left(\frac{\nu-\nu_0}{\nu_0}\right)^t I^{psf}_{\nu}
%\end{equation}
%where $t=1:n_{taylor}$ where $n_{taylor}$ is the order of the Taylor expansion
%and $h_s$ is an inverted paraboloid function whose width is given by $s$. 
%The Minor Cycle solves for $C(t,s)$ from which spectral index and curvature images are
%calculated. 
This algorithm is suited for  extended emission and features with non-linear
spectra described by a power law of varying index across the observing band.
%($\alpha \neq constant$).

Some direction-dependant effects in $K^{sky}_{ij}$
(e.g. effect of the Primary Beam) are also frequency dependant.
Therefore the {\it spectral PSFs} and {\it dirty images} used 
in the Minor Cycle can be computed as another
generalization of Eq.~\ref{Eq:PBDIRTY}.
\begin{equation}
I^{\{dirty,\atop psf\}}_t = [I^{wt}_{dd}]^{-1}[I^{wt}_{gc}]^{-1} \sum\limits_{\nu} [\Fd\G^{gc}\G^{dd^\dag}_{\nu}\Sd\W^{im}_{\nu,t}] V^{\{corr,1\}}_{\nu}
\label{MFDD}
\end{equation}
where 
\begin{equation}
\W^{im}_{\nu,t} = \W^{im} \left((\nu-\nu_0)/\nu_0\right)^t
\label{SpWts}
\end{equation}
The {\it weight} image 
describes the noise variation across the image due to imaging weights and 
frequency dependant $K^{sky}_{ij}$ and is given by
\begin{equation}
I^{wt}_{dd} = \sum\limits_{\nu} [\Fd\G^{dd^\dag}_{\nu} \W^{im}_{\nu,t=0} \G^{dd}_\nu\F]
\end{equation}

%
%====================================================================
%
\subsection{Full polarisation Calibration and Imaging}\label{Full Stokes}
The preceeding sections have dealt with the calibration and imaging of only one
correlation pair $pp$ from a single feed. 
This section deals with the full-polarisation calibration of a pair of 
potentially imperfect orthogonal feeds,
and the imaging of all four Stokes parameters.

\subsubsection{Full-Stokes Calibration}
Each baseline measures the product of $K^{vis}_{ij}=J^{vis}_i \otimes {J^{vis}_j}^*$ 
with the true coherence vector seen by that baseline. Eq.~\ref{Eq:VIJ-FT} 
becomes 
\begin{equation}
\vec{V}^{obs}_{4n\times 1} = [K^{vis}_{4n\times 4n}] \vec{V}^{model}_{4n\times 1}
\label{eqnSELFCALPOL}
\end{equation}
and the elements of $K^{vis}_{ij}$ are computed as described in section \ref{CALIBRATION}. 
For a source with known polarisation characteristics,
the true coherence vector is known (constant $\times$ [1,0,0,1] for circular feeds and an
unpolarised source) and one can form a
system of linear equations with the elements of $K^{vis}_{ij}$ as unknowns. 
For a single baseline, there are up to 10 degrees of freedom and 4 equations \cite{HBS2}.
%and this can be solved in
%isolation only by adding in assumptions of relations between the terms. 
However, with an a-priori source model, measurements from all baselines 
provide enough constraints to uniquely factor the 
baseline-based $K^{vis}_{ij}$ matrices into antenna-based Jones matrices 
($4 \times n_{ant}(n_{ant}-1)/2$ equations and $4\times n_{ant}$ unknowns).
In its most general form, the elements of $J^{vis}_i$ can be computed by minimizing
$\chi^2 = \sum_{ij} | \vec{V}^{obs}_{ij} - [J^{vis}_i \otimes {J^{vis}_j}^*]\vec{V}^{m}_{ij}|^2$
w.r. to the antenna based $J^{vis}_i$.

In existing software packages, polarisation calibration is usually done in stages.
First, only the diagonal elements of the Jones matrices are solved for, assuming zero leakage
between the orthogonal feeds. Solutions are then applied and only off-diagonal terms
are solved for. Another method of solving for antenna based gains and leakages
from only parallel-hand correlations $pp,qq$ is described in \cite{LEAKYCAL}.
%and is equivalent to the eigen-decomposition of a rank 2 matrix.
The effects of depolarisation cannot be factored into Jones matrices and 
a baseline-based calibration is sometimes carried out by artificially imposing 
constraints between the elements of $K^{vis}_{ij}$.

\subsubsection{Full-Stokes Imaging}
The Stokes vector for polarised sky brightness $\vec{I}^{stokes} = \{I,Q,U,V\}$ is related to
the vector of images corresponding to the correlations $\{pp,pq,qp,qq\}$ as
\begin{equation}
\vec{I}^{sky}_{4m \times 1} = [\mathcal{S}_p]_{4m\times 4m} \vec{I}^{stokes}_{4m\times 1}
\label{eqnCorrToStokes}
\end{equation}
where $\mathcal{S}_p$ holds a $4\times 4$ linear operator per image pixel \cite{HBS1}.
%From eqn \ref{Eq:VIJ3}, the full-Stokes measurement equation is
%\begin{equation}
%\vec{V}^{corr}_{4n\times 1} = [S_{4n\times 4m}] [F_{4m\times 4m}] [K^{sky}_{4m\times 4m}] [{\mathcal{S}_p}] \vec{I}^{Stokes}_{4m\times 1}
%%%\vec{V}^{corr}_{cn\times 1} = [S_{cn\times cm}] [F_{cm\times cm}] [K^{sky}_{cm\times cm}] [{\mathcal{S}_p}_{cm \times cm}] \vec{I}^{Stokes}_{cm\times 1}
%\end{equation}
%where $[{\mathcal{S}_p}]$ holds a $4\times 4$ linear operator per image pixel and 
%$[K^{sky}]$ is known in the correlation basis.
A full-Stokes deconvolution differs from standard methods in the computation
of {\it dirty} images and the {\it Minor cycle}.
%where the interdependence between coherence images (elements of $\vec{I}$)
%needs to be taken into account.
%Since the images in the correlation basis (elements of $\vec{I}$) are not linearly-independant
The Stokes vector of dirty images $\vec{I}^{dirty,Stokes}$ is computed by applying
Eq.~\ref{eqnCorrToStokes} to the set of dirty images in the correlation basis
$\vec{I}^{\{dirty,corr\}}$ given by Eqs.~\ref{DIRTYANDPSF} or \ref{Eq:PBDIRTY}.
The different Stokes parameters are considered to be 
linearly independent and deconvolution minor cycles are performed separately on each
Stokes image. For compact sources, position constraints are sometimes applied 
across Stokes parameters based on the locations of peak residuals
of the Stokes I image.
\cite{MEMFULLPOL} describes an algorithm that applies the
constraint of $I^2 \ge Q^2 + U^2 + V^2$ during deconvolution.

\section{Conclusion}

We have presented a complete mathematical framework for describing many of the major calibration and imaging algorithms used in radio interferometry. This framework can be used for three purposes:
%\begin{itemize}
%\item Elucidating the fundamental assumptions and details of algorithms,
%\item Isolating the mathematical structure so that standard libraries can be used,
%\item Allowing both generalization and specialization to generate new algorithms.
%\end{itemize}
(a) Elucidating the fundamental assumptions and details of algorithms,
(b) Isolating the mathematical structure so that standard libraries can be used,
and (c)  Allowing both generalization and specialization to generate new algorithms.

The computing and software issues connected with the use of this framework are substantial, especially given the large data volumes and processing loads being contemplated for new radio telescopes. We will discuss these issues further in a subsequent paper.
We note that this framework can also be used to address other
algorithms not discussed here. These include the {\it peeling}
technique for direction-dependent calibration, the problem of {\it
  ionospheric calibration} as a direction-dependent effect, and the
excision of radio-frequency-interference from measured visibility
data.

% or


\appendices
%
%====================================================================
%
%\section{Title of Appendix 1}

%\section{This is Appendix 2}

\section*{Acknowledgment}

The authors would like to thank Kumar Golap and Rajaram Nityananda for helpful discussions.
We would also like to thank Richard Thompson and the referees for their careful
reading of this paper, and helpful comments and suggestions.

Our work resides in two principal software packages : the widely
available CASA project (previously AIPS++) from 
NRAO,
%NRAO\footnote{(The National Radio Astronomy Observatory (NRAO) is a facility
%		of the National Science Foundation operated under cooperative 
%		agreement by Associated Universities, Inc.)},
and the software
being developed for ASKAP.  The AIPS++/CASA synthesis code framework for
algorithm development and implementation is now quite mature, 
having been initially developed about ten years ago
\cite{CornwellWieringa1997}, though with many additions. The ASKAP
radio interferometry code is substantially new, though some of the low
level libraries are common with CASA. We thank the many people who have
contributed to both packages.

\ifCLASSOPTIONcaptionsoff
  \newpage
\fi



\bibliographystyle{IEEEtran}
\bibliography{aph_ieeemeqn}
\end{document}